# DATABASE AND CHANNEL ACCESS ON THE VEPP-4 CONTROL SYSTEM

D. Filimonov, S. Karnaev, B. Levichev, S. Smirnov, A.Veklov,
Budker Institute of Nuclear Physics, Novosibirsk

Abstract

The VEPP-4 control system was designed almost 20 years ago as CAMAC-based distributed control system [1]. Today the main peculiarity of the upgraded VEPP-4 [2] control system is the employment of obsolete CAMAC-embedded 24-bit Odrenok computers in the middle (process) level.

The communication software is developed to provide an effective interaction between user (operator) applications on the high level and executive programs in the middle level. The communication software is Linux-based and includes the Application Server and Device Servers. The Application Server provides channel and database access for user applications. The Device Servers provide interaction with the hardware. The backend database server is based on PostgreSQL.

This paper describes the structure of the VEPP-4 control system database and data/channel access procedures.

## 1 INTRODUCTION

The VEPP-4 database for the control programs in the Odrenok computers is distributed in separate files according to the VEPP-4 complex subsystems. These files include information about electronics, control and measuring channels. Data format is 24-bit ICL word used in Odrenok. A few types of custom text editors are used for database viewing and editing.

The grave imperfection of Odrenok-based database was a lack of the possibility to log large numbers of data from equipment, like the vacuum gauge, temperature, magnet current, etc in a uniform way. The data acquisition programs running on Odrenoks support limited collection of the data.

The first attempts to develop a unified database in the VEPP-4 control system were made in the beginning of the 90's, but were not successful because of the low bandwidth connection among Odernok computers and the PCs. The real possibility to adopt professional database system for the VEPP-4 occurs with implementation Ethernet and using PCs under Linux [2].

The first goal of the new database is to provide an access to the data about all control and measuring channels of present control system for the new applications running in the PCs. There are data converters to produce special data files for initial program loading and running of the applications in each Odrenok.

The next goal is to provide channel access between new applications running in the PCs and the executive programs in Odrenoks.

The scalability of the database system should be flexible for modification in the case of addition of new accelerator control subsystems.

## 2 DATABASE ARCHITECTURE

### 2.1 General description

The VEPP-4 database includes two parts. The first part is a static data provided via PostgreSQL server. This part consists of relational tables with all necessary entities for control.

The second part includes the database servers providing data access and data archiving. The main part of the data access system is Application Server (AS). The AS is a dispatcher for all the requests from clients to database, control and measuring channels. If the request is to put/get data in/from control/measuring channel, the AS transmits the request to the corresponding Device Server running in the computer which is connected to the corresponding electronics. If the request is directed to executive programs working on Odrenok computer it handled via Odrenok Server.

A simplified diagram of the VEPP-4 database is shown in Fig.1.

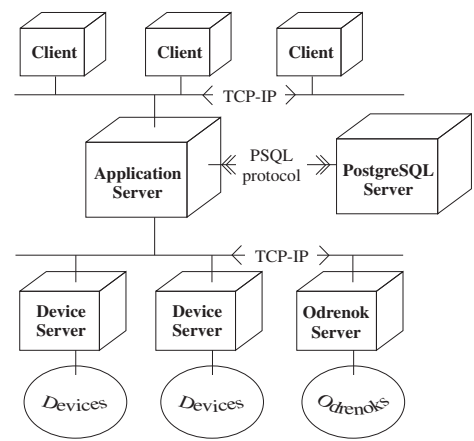

Figure 1: The VEPP-4 database diagram.

The peculiar properties of the database system are:
- Methods of handling a query are encapsulated into the query. This makes query handling independent of requests and provides enlarging set of queries easily.
- The use of the benefits of QT v.3 [3], the most significant of them are:
  - Signal-slot architecture [4], [5]
  - Asynchronous sockets
  - Unified access to RDBMS

### 2.2 Tables

The static database based on PostgreSQL includes a set of tables with description of the electronics, control/measurement channels, objects of control, stored operation modes, etc. The main database tables are:
- **Objects.** This table includes a description of all physical and logical elements and objects in the VEPP-4 facility such as magnets, power supplies, control elements, etc. All objects are involved into the hierarchical structure.
- **ObjectParameters.** This table includes different attributes and characteristics of the objects from the "Objects" table, for example: coil resistance and field/current ratio for the bending magnet.
- **Modules.** It is a description of all the control and measuring electronic modules and units.
- **Crates.** This table includes a description of the physical or logical groups of electronic modules: CAMAC crates, serial buses, etc. Tables "Modules" and "Crates" provide a convenient way for addressing to modules in programs.
- **Channels.** Each record of this table (channel) includes all necessary data for manipulation or monitoring particular physical parameters: code/units ratio, address, limits, etc. There are about 2000 control and measuring channels on the VEPP-4 not including temperature, vacuum, and beam pick-up diagnostics. There are separate tables of channels for these systems.
- **OperationModes.** This table includes a description of the stored operation modes. Each operation mode stored in a separate file according to a specified template. Each template for creating an operation mode includes a list of channels.

The number of the different table types in the VEPP-4 database is about twenty.

### 2.3 Object oriented scheme

An application should send a specified request (query) to get an information from the VEPP-4 database. Application Server can handle a certain set of the object related query types: get/put crate, get/put channel, etc. The VEPP-4 database is an assembly of different type objects for clients.

The VEPP-4 database object conception is shown in Fig.2.

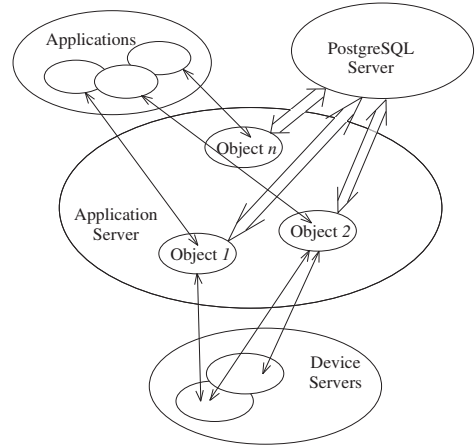

Figure 2: The VEPP-4 database object representation.

The Application Server (AS) creates corresponding object for handling each incoming query. The objects encapsulate data from postgres server and methods for the data processing. The objects interact with applications, device servers and postgres server via established separate connections. After the query is performed corresponding object remains in the AS during a certain time for the possible demand in the next period.

## 3 APPLICATION SERVER

The Application Server is the main part of the VEPP-4 database system. The AS block diagram is represented in Figure 3.

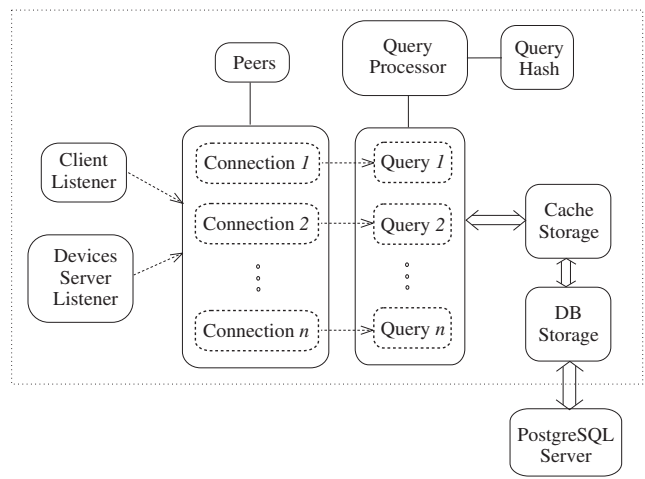

Figure 3: Application Server diagram.

The AS consists of several functional components. The following components are the same for AS, Device Servers and client applications:
- **Query Hash**. All requests being handled are stored in the hash and each part of the program references the query through its hash key only. This allows handling cases when referenced query was removed to be safe with a little penalty performance. But stability is much more important in this case.
- **Query Processor**. It contains common logic of request processing independent of queries and handles queues of queries being processed.
- **Queries**. Queries are a set of requests containing methods of processing in addition to data.
- **Connections.** Connections are a set of TCP connections between AS and Clients, and between AS and Device Servers.

Components specific to the Application Server are:
- **Storage.** It is an abstract class defining set of the operations for saving/loading persistent objects.
- **DB Storage.** It is a realization of the **Storage** implementing database storage.
- **Cache Storage.** It is a realization of the **Storage** catching requests to the **DB Storage** for a better performance.
- **Peers.** It is a hash of **Connections**; it is needed for request delivery.

## 4 DEVICE SERVERS

The Device Server (DS) is the part of database system corresponding to the data transfer between AS and devices. The DS makes front-end processing.

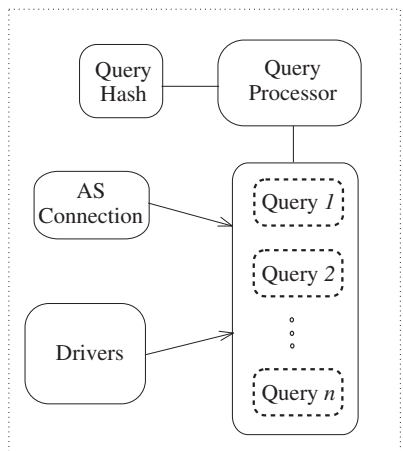

Figure 4: Device Server diagram.

The specific components of the DSs are **Device Drivers** and **AS connection**. Driver is the logic of the device management. The AS connection is a TCP connection between DS and AS.

In the case of Odrenok Server the driver provides the query transfer from the PC to the Odrenok executive programs.

## 5 CLIENTS

The client applications provide visualization and monitoring, operator interface for control, data storing, etc.

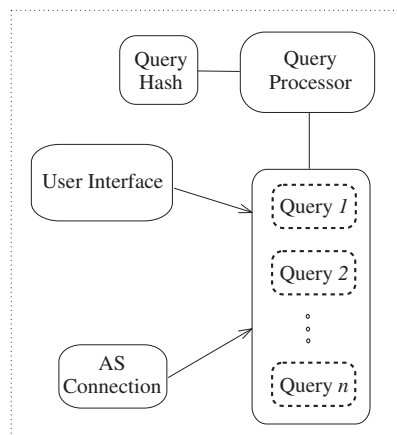

Figure 5: Application diagram.

The specific components of the Clients are:
- **User Interface**. It is a logic implementing client specific behavior. It may communicate to an operator, or it may be an offline data-manipulating program.
- **AS connection** is a TCP connection between an application and AS.

## 6 CONCLUSION

Now the VEPP-4 database system referred above has been implemented. The first stage provides the conversion of all VEPP-4 control system data to the postgres database. This stage provides loading and running Odrenok executive programs.

The second goal to provide an accelerator operator view and control via PC applications will be achieve in the near future.